\documentclass[12pt]{article}

\usepackage{scicite}
\usepackage{times}
\usepackage{setspace}
\usepackage{graphicx}

\newenvironment{sciabstract}{
\begin{quote} \bf}
{\end{quote}}

\newcounter{lastnote}
\newenvironment{scilastnote}{%
\setcounter{lastnote}{\value{enumiv}}%
\addtocounter{lastnote}{+1}%
\begin{list}%
{\arabic{lastnote}.} {\setlength{\leftmargin}{.22in}}
{\setlength{\labelsep}{.5em}}} {\end{list}}

\title{Quantum-State Controlled Chemical Reactions of Ultracold KRb Molecules}

\author{S. Ospelkaus,$^{1\ast}$ K.-K. Ni,$^{1\ast}$ D. Wang,$^{1}$ M. H. G. de Miranda,$^{1}$ B. Neyenhuis,$^{1}$\\ G. Qu\'em\'ener,$^{1}$ P. S. Julienne,$^{2}$ J. L. Bohn,$^{1}$ D. S. Jin,$^{1\dagger}$ J. Ye$^{1\dagger}$\\
\\
\normalsize{$^{1}$JILA, NIST and University of Colorado,}\\
\normalsize{Department of Physics, University of Colorado, Boulder}\\
\normalsize{Boulder, CO 80309-0440, USA}\\
\normalsize{$^{2}$Joint Quantum Institute, NIST and University of Maryland, }\\
\normalsize{Gaithersburg, MD 20899-8423, USA}\\
\\
\normalsize{$^\ast$These authors contributed equally to this work; }\\
\normalsize{$^\dagger$To whom correspondence should be addressed; }\\
\normalsize{E-mail:  Jin@jilau1.colorado.edu; Ye@jila.colorado.edu}
}
\begin{document}

\baselineskip24pt

\maketitle

\begin{sciabstract}
How does a chemical reaction proceed at ultralow temperatures? Can simple quantum mechanical rules such as quantum statistics, single scattering partial waves, and quantum threshold laws provide a clear understanding for the molecular reactivity under a vanishing collision energy? Starting with an optically trapped near quantum degenerate gas of polar $^{40}$K$^{87}$Rb molecules prepared in their absolute ground state, we report experimental evidence for exothermic atom-exchange chemical reactions. When these fermionic molecules are prepared in a single quantum state
at a temperature of a few hundreds of nanoKelvins, we observe $p$-wave-dominated quantum threshold collisions arising from tunneling through an angular momentum barrier followed by a near-unity probability short-range chemical reaction. When these molecules are prepared in two different
internal states or when molecules and atoms are brought together, the reaction rates are enhanced by a factor of 10 to 100 due to
$s$-wave scattering, which does not have a centrifugal barrier. The measured rates agree with predicted universal loss rates related to
the two-body van der Waals length.

\end{sciabstract}

Scientific interests in precisely understanding the fundamental aspects of chemical reactions and controlling their dynamic processes have stimulated pioneering work on molecular beams to study state-to-state reactions using molecular alignment, velocity selections, and angle-resolved measurement~\cite{Lee,Herschbach,Zare,Welge}. However, substantial motional energies remained in earlier work and thermal statistical averages were a necessary ingredient. By preparing a molecular ensemble in the quantum regime, we expect to develop fundamental insights into how chemical reaction processes may be precisely guided by quantum mechanics. Reaction dynamics at vanishingly low energies remain a fascinating and yet unexplored scientific realm~\cite{Carr2009a}. Here both the internal and external molecular degrees of freedom are prepared quantum mechanically, and each step of a complex reaction may be analyzed based on single quantum states and single reaction channels.

That chemical reactions could occur at ultralow temperatures seems at first glance counter-intuitive. However, ultracold collisions,
where particles scatter only in the lowest angular momentum partial wave, are governed by quantum statistics and quantum threshold behaviors
described by the Bethe--Wigner laws~\cite{Bethe,Wigner,Sadeghpour2000}. In this regime, particles are represented by their de Broglie wavelength, which increases with reduced temperatures. This wave nature of particles replaces our intuitive and classical picture of collisions. The wave manifestation of tunneling through reaction or angular momentum barriers may play a dominant role in dynamics and scattering resonances can have dramatic effects on reactions~\cite{Balakrishnan2001}.  In addition, any barrierless chemical reactions will always take place when two reactants are sufficiently close together \cite{Hutson2007}. In this case, chemical reaction rates will be determined to a large extent by collisional properties at large intermolecular separations, and thus by how the two partners approach each other. Once their separation reaches a
characteristic length scale ($\sim$10~$a_0$, with $a_0=0.53\cdot 10^{-10}$~m), a chemical reaction happens with a near unity probability.  Therefore, chemical reactions can be surprisingly efficient even at ultracold temperatures. Indeed, this model for barrierless reactions predicts loss rates that are universal in the sense that they do not depend on the details of the short-range interactions, but instead can be estimated using only knowledge of the long-range interactions~\cite{Julienne09a}.

Like the case of collisions of ultracold atoms, the study of ultracold chemical reactions will play a fundamental role in
advancing the field of molecular quantum gases. For example, understanding and manipulating collisions of
atoms at ultralow temperatures ($<$ 1$\mu$K) have been crucial for the realization of quantum degenerate gases~\cite{BEC, BEC2, DFG},
Fermi superfluids that provide opportunities to explore the underlying connection between superconductivity and Bose-Einstein condensation~\cite{CindyThesis}, neutral atom-based systems for quantum information science \cite{Jaksch1998, Gaetan2009, Urban2009},
and strongly correlated quantum gases \cite{Bloch2008, Kinoshita2006}. Ultracold molecules undergo a more diverse set of
collisional processes, with distinct inelastic collision mechanisms arising from chemical reactions, in addition to the
traditional state-changing collisions seen with ultracold atoms and highly vibrationally excited molecules\cite{Hudson2008}.
Furthermore, polar molecules possess anisotropic and long-range dipolar interactions that can be precisely controlled with external
electric fields, with rich prospects of collisional resonances~\cite{Bohn2002,Bohn2005}. Experimental investigation of molecular collisions is essential for future applications including studying anisotropic and collective behavior in quantum gases \cite{Lahaye2009a, Yi2000}, modeling new
quantum phases and exotic many-body physics \cite{zollerCM}, implementing schemes for quantum information \cite{Andre2006}, and developing
tools for precision physical and chemical measurements.

We focus here on the study of ultracold collisions, including chemical reactions, of $^{40}$K$^{87}$Rb molecules, which we prepare in their lowest electronic, vibrational, rotational, and hyperfine energy state at a high phase-space density \cite{Ni2008, Ospelkaus2009b}.
We find clear evidence of the essential role that quantum statistics and quantum threshold laws play in determining the rates
of inelastic collisions. Our experimental observations confirm a universal loss mechanism.

A prerequisite for exploring ultracold chemical reactions is a gaseous molecular sample that is sufficiently dense, ultracold, and
suitable for precise control of specific quantum states~\cite{Carr2009a}. The starting point for this work is an ultracold trapped gas of fermionic
$^{40}$K$^{87}$Rb molecules prepared in a single hyperfine level of the ro-vibronic ground state ($N=0,\, v=0$ of
$X^1\Sigma^+$)~\cite{Ni2008, Ospelkaus2009b}. The optical trap depth is $\sim k_B \cdot $10~$\mu$K, where $k_B$ is
Boltzmann's constant. The molecules are produced using a single step of two-photon Raman transfer from extremely weakly bound molecules
at a magnetic field of $\sim$545.9~G. The coherent transfer is efficient and does not heat the gas, resulting in a gas of
ro-vibronic ground-state molecules with an average number density of $10^{11}$ to $10^{12}$~cm$^{-3}$ and a translational temperature of a few
hundreds of nanoKelvin. At this ultralow temperature, even the tiny molecular hyperfine-state energy splittings are much larger than the
translational energy. Manipulation of the hyperfine states hence becomes extremely important and relevant in exploring
possible collision channels.  In addition, complete control over the internal quantum state of the molecules permits direct observation
of the role of quantum statistics in determining the molecular interactions.

\begin{figure}[t]
\begin{center}
\includegraphics[]{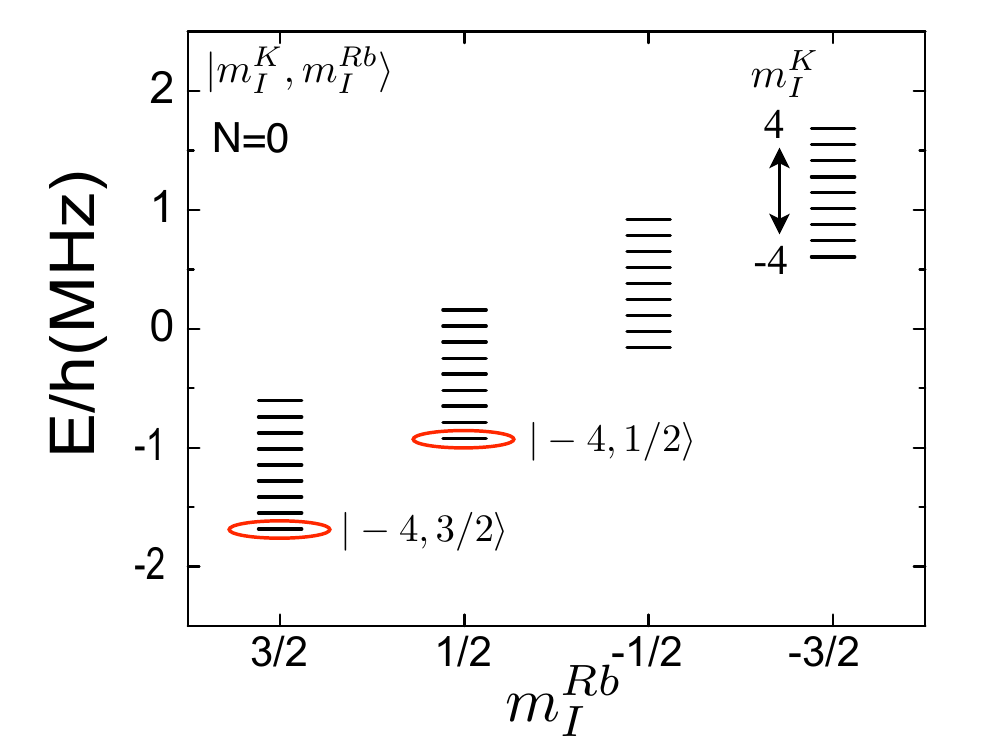}
\caption{ \doublespacing Hyperfine structure of ro-vibronic ground-state $^{40}$K$^{87}$Rb
molecules at 545.9 G. We label the 36 nuclear spin states by their spin projections,
$m_I^{Rb}$ and $m_I^{K}$. The energy spacing between hyperfine states is $\sim h\cdot$130 kHz for
$|\Delta m_I^K|=1$ and $\sim h\cdot$760 kHz for $|\Delta m_I^{Rb}|=1$.
By comparison, at a temperature of 300 nK, the molecules' thermal energy
is equivalent to $\sim h\cdot$6 kHz, which is more than an order of magnitude smaller than the
spin flip energy. In our experiments, molecules are
prepared in either a single state or in a mixture of
$|-4,1/2\rangle$ and the lowest energy state $|-4,3/2\rangle$ (open ellipses).
 \label{hyperfine}}
\end{center}
\vspace{-0.05\textwidth}
\end{figure}

A precise measurement of the hyperfine structure and the manipulation of individual hyperfine state populations in the $X^1\Sigma^+$ ground state of $^{40}$K$^{87}$Rb were reported in \cite{Ospelkaus2009b}. The $X^1\Sigma^+$ state has zero total electronic angular momentum,
so that the hyperfine structure is basically the Zeeman effect of the nuclear spins $I^K$=4 and $I^{Rb}$=3/2~\cite{Aldegunde08, Julienne09a} at
the applied magnetic field.  The hyperfine structure is depicted in Fig.~\ref{hyperfine}, where a total of 36 states are
labeled by their projections of the individual nuclear spins, $m_I^{Rb}$ and $m_I^{K}$. For the current study, we produce molecules either in a single spin state ($|m_I^{K},m_I^{Rb}\rangle$) or in a mixture of two spin states.  The hyperfine states used are an excited state $|-4,\,1/2\rangle$ and the lowest energy spin state $|-4,\,3/2\rangle$; these are marked by the two ellipses in Fig.~\ref{hyperfine}. The $|-4,\,1/2\rangle$ state is populated directly by the two-photon Raman transfer starting from the weakly bound molecules~\cite{Ni2008, Ospelkaus2009b}. The spin state of these molecules can be further manipulated coherently. For example, the entire $|-4,\,1/2\rangle$ population can be transferred into the lowest hyperfine state, $|-4,\,3/2\rangle$, using two successive $\pi-$pulses through a rotationally excited $N=1$
intermediate level. The $N=1$ level has strong nuclear electric quadrupole interactions that couple rotations with nuclear spins,
enabling nuclear spin flips~\cite{Ospelkaus2009b,Aldegunde08}. We can probe molecules in any particular hyperfine state by reversing the entire
transfer process and putting the population back into the initial weakly bound state. We then use high signal-to-noise-ratio absorption
imaging to measure the molecular gas number and temperature.

The ability to control molecular internal states including the electronic, vibrational, rotational, and nuclear spin degrees of freedom, and in
particular the possibility of preparing them in the lowest energy state, allows us to probe inelastic collisions in a way that limits
unwanted loss mechanisms and reduces ambiguities in the identification of possible chemical reaction channels. This is crucial
because it is difficult in the ultracold gas experiment to find probes for the direct observation of reaction products.
In Table \ref{potential}, we consider the binding energies for various types of molecules made from different
combinations of $^{40}$K and $^{87}$Rb atoms. These energy estimates allow us to assess whether a specific two-body reaction process is
endothermic or exothermic. We note that all the results reported here were obtained in the absence of any external
electric field, and hence the effective molecular dipole moment in the lab frame is zero.

\begin{table}[ht]
\begin{center}
\caption{\label{potential} Summary of the relevant molecular energetics involved in possible chemical reactions.  The binding energies are given
with respect to the threshold energy for free atoms in the absence of a magnetic field. The $^{87}$Rb$_2$ and $^{40}$K$_2$ binding energies include isotope shifts from the data in the respective references. The trimer binding energies are unknown. Calculations of trimer
binding energies are needed and will be important for future experiments on any bi-alkali species.}
\begin{tabular}{|c|c|c|c|c|}\hline
\centering Molecules& $v=0$ binding energy ($D_0$) & reference \\\hline
\centering $^{87}$Rb$_2$ & 3965.8(4)~cm$^{-1}$& \cite{Amiot90} \\\hline
\centering $^{40}$K$^{87}$Rb& 4180.417~cm$^{-1}$ & \cite{Ni2008} \\\hline
\centering $^{40}$K$_2$&  4405.389(4)~cm$^{-1}$& \cite{Falke2006} \\\hline
 \centering K$_2$Rb& unknown &  \\\hline
     \centering KRb$_2$&  unknown &  \\\hline
\end{tabular}
\end{center}
\end{table}

To probe the quantum nature of ultracold molecular collisions and chemical reactions, we begin by preparing the KRb molecules in a single
nuclear spin state of the ro-vibronic ground state. All unpaired atoms that remain after the initial stage of the molecular creation
process are selectively removed from the optical trap via resonant light scattering~\cite{removal}.  From an argument
based on the energetics summarized in Table~\ref{potential}, the molecule-molecule collisions have a possible exothermic chemical
reaction, namely KRb + KRb$\rightarrow$K$_2$ + Rb$_2$, which releases $\sim$10 cm$^{-1}$ of kinetic energy.
The reactions KRb + KRb$\rightarrow$K$_2$Rb + Rb and KRb + KRb$\rightarrow$KRb$_2$ + K could also be exothermic.
All of these reactions require breaking and making molecular bonds. If the KRb molecules are prepared in an excited hyperfine state,
spin relaxation to a lower hyperfine state provides an additional inelastic scattering mechanism.

The quantum statistics of the molecules plays an essential role in collisions at a temperature of a few hundred
nanoKelvin, where collisions with large-impact parameters and correspondingly large centrifugal barriers are frozen out and the collisions are typically dominated by a single partial wave with orbital angular momentum quantum number $L$=0 ($s$-wave) or $L$=1 ($p$-wave). Our KRb molecules are fermions and therefore the total wave function describing a KRb+KRb collision is antisymmetric with respect to molecular exchange. For
spin-polarized molecules all prepared in exactly the same internal quantum state, the $p$-wave is the lowest energy symmetry-allowed
collision channel. The height of the centrifugal barrier for the $L$=1 KRb-KRb collisions is $k_B\cdot 24\,\mu$K
\cite{SvetlanaC6, KRbC6}. This barrier height is over an order of magnitude larger than $k_B T$, where $T$ is the translational temperature of the
molecular gas. Thus, collisions of spin-polarized molecules are expected to proceed predominately via tunneling through the
$p$-wave barrier. Note that $T$ is greater than 1.4 times the Fermi temperature.  If two molecules make it through the barrier to
short range, chemical reactions or hyperfine state-changing collisions can take place, leading to a loss of the entrance channel population. We note that even a single nuclear spin flip corresponds to a released quantity of
energy that is above the trap depth, and would contribute to
loss of trapped molecules.

\begin{figure}[t]
\begin{center}
\includegraphics[width=\columnwidth]{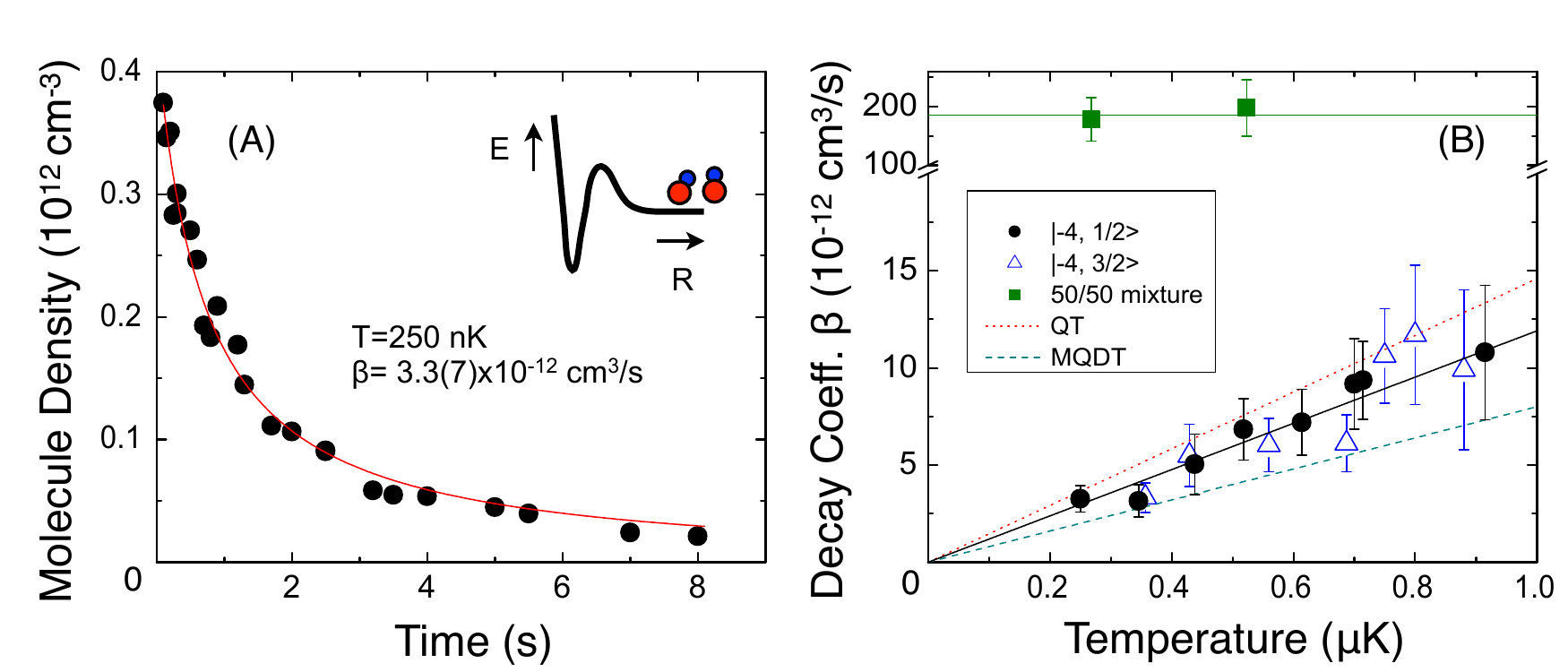}
 \caption{\doublespacing Inelastic collisions between spin-polarized (indistinguishable) or
different spin-state (distinguishable)
fermionic molecules in the ro-vibronic ground state of $^{40}$K$^{87}$Rb.
 (A)  Sample data showing the time dependence of the molecule number
density. Here the molecules are prepared in a single hyperfine state, $|-4, 1/2\rangle$,
and the molecular
 density decays slowly with a rate coefficient of $3.3(7)\cdot 10^{-12}$ cm$^{3}$/s at $T=250$ nK.
 (B) Loss rate coefficient vs temperature. The loss rate increases
 linearly with temperature for spin-polarized molecules, which collide via $p$-wave (inset to A)
 at low temperature. Data were taken for
 molecules prepared in either $|-4, 1/2\rangle$ (closed circles) or the lowest
  energy state $|-4, 3/2\rangle$ (open triangles). A linear fit (solid line) to the $|-4, 1/2\rangle$ data
  yields the temperature-dependent loss rate to be
  1.2(3)$\cdot 10^{-5}$ cm$^3$/s/Kelvin.
  For the $|-4, 3/2\rangle$ case, where the collisional loss can only be due
  to chemically reactive scattering, the loss rate is similar. The dotted and
  dashed lines are theoretical predictions
  from the QT model and MQDT, respectively.
      When the molecules are prepared in a
  mixture of  the $|-4, 1/2\rangle$ and $|-4, 3/2\rangle$
   states (filled squares), we observe a temperature-independent decay rate that is 10 to 100
   times higher than for the spin-polarized case. The error bars represent 1 SD of the decay rate coefficients arising from fluctuations of the  molecular density, temperature, and fitting uncertainty of the two-body loss curves.
   \label{mmDecay}}
\end{center}
\vspace{-0.05\textwidth}
\end{figure}

The quantum nature of the collisions can be seen in the temperature dependence of loss rates. The Bethe-Wigner threshold law predicts that the $p$-wave inelastic/reactive collision rate should be linear in temperature ($\propto T$). To look for this behavior, we first prepared spin-polarized molecules in the single hyperfine state $|-4, 1/2\rangle$ for various $T$ ranging from 200 to 900 nK\cite{temp}.
The temperature is measured from the expansion energy of the molecular gas after releasing it from the optical trap. For each initial
temperature, we observed the time-dependent molecular loss \cite{blackbody} and extracted a two-body loss rate $\beta$ (which is 2 times the collisional event rate) by fitting the measured decay of the molecular gas density $n$ vs. time
$t$ (Fig. \ref{mmDecay}(A)) to 
\begin{eqnarray}
 \frac{dn}{dt}=-\beta n^2-\alpha n.
\end{eqnarray}
Here, the first term accounts for number loss and the measured $\beta$ can be compared to theoretical predictions. The second term accounts
for density change due to heating of the trapped gas during the measurement. Within a single measurement, we observe an increase in temperature that is at most 30\%.  In the analysis for each data set, we fit the measured temperature to a linear heating rate and obtain a constant slope $c$.  In Eq. 1, we then use $\alpha=\frac{3}{2}\frac{c}{T+c\,t}$, where $T$ is the initial temperature.
At our lowest temperature of 250 nK, the heating was 7(1) nK/s and $\beta=3.3(7)\cdot 10^{-12}$cm$^{3}$/s (Fig.~\ref{mmDecay}(A)). The measured dependence of $\beta$ vs. $T$ is summarized in Fig.~\ref{mmDecay}(B) (closed  circles). Here, we fit the data to a power law $\beta(T)\propto T^L$ and find that $L=1.1(2)$, which agrees with the predicted $p$-wave threshold law. This result demonstrates that indistinguishable
$^{40}$K$^{87}$Rb molecules at ultralow temperatures collide via tunneling through a $p$-wave barrier followed by an inelastic
collision in the short range. A linear fit to the data ($L=1$) yields a slope of the decay rate coefficient of $1.2(3)\cdot
10^{-5}$ cm$^3$/s/Kelvin.

We repeated this measurement for molecules in the lowest hyperfine state $|-4, 3/2\rangle$ (open triangles in Fig.~\ref{mmDecay}(B)).
The data again show $\beta \propto T$ with a slope of $1.1(3)\cdot 10^{-5}$ cm$^3$/s/Kelvin, similar to that measured for molecules in
the $|-4, 1/2\rangle$ state. However, in the case of $|-4, 3/2\rangle$ molecules, hyperfine state-changing collisions are no
longer possible and the only possible loss channels are the chemical reactions discussed above. Thus, we find that the rate of
chemical reactions is determined by the $p$-wave angular momentum barrier and the chemical reaction barrier must be
below the collision energy. This suggests that these reactions are barrierless and can thus occur freely at ultralow
temperatures. Meanwhile, the fact that the same loss rate is observed for both $|-4, 1/2\rangle$ and $|-4, 3/2\rangle$ state
molecules suggests that chemical reactions dominate the loss in these ground-state molecular collisions.

To understand the loss rates, we use two models: one is a simple quantum threshold model (QT)
and the second is a model that uses the formalism of multi-channel quantum defect theory (MQDT). In the QT model, the
loss rate for collisions with energy equal to or above the height of the $p$-wave barrier
is determined by the Langevin capture rate~\cite{Langevin05}, which assumes that the probability for chemical reactions/hyperfine state-changing
collisions is unity. For all collision energies below the height of the $p$-wave barrier,
we assume in this model that the loss follows the Bethe--Wigner threshold laws~\cite{Bethe,Wigner}.
Using this assumption, we obtain a simple analytical expression for the $p$-wave loss rate coefficient
of two indistinguishable molecules, which scales linearly with $T$~\cite{Quemener2009a}:

\begin{eqnarray}
\beta=\frac{\pi}{4} \, \left(\frac{3^{17} \, \mu^3 \,
C_6^3}{\hbar^{10}}\right)^{1/4} \, k_B T,\,
\end{eqnarray}

where $\mu$ is the reduced mass and $\hbar$ is the Planck's constant divided by $2\pi$. Using a van der Waals dispersion
coefficient of $C_6$=16130 a.~u. for KRb-KRb with a $\pm 10 \%$ uncertainty~\cite{SvetlanaC6,KRbC6},
the slope of the rate coefficient is predicted to be $1.5(1)\cdot 10^{-5}$ cm$^{3}$/s/Kelvin, which agrees well with the experimental measurement.

In the second model, the loss rate coefficient is found directly by calculating the quantum tunneling rate through the $p$-wave barrier \cite{Idziaszek}. This calculation gives $\beta=0.8(1)\cdot 10^{-5}$ cm$^3/$s/Kelvin, which agrees with the experiment within mutual uncertainties.  This $\beta$ can also be derived analytically from the properties of the long-range potential to give $\beta = (11.48 \bar{a})^3$ (k$_B T/h$), where $\bar{a}=0.4778 (2\mu C_6/\hbar^2)^{1/4} = 6.3$ nm is the characteristic length of the van der Waals potential~\cite{Julienne09a}.  This fully quantum calculation can be put in the same form as the QT model Eq. (2), and gives a $\beta$
that is smaller by a factor of 0.528. With these simple theories, the agreement with our molecule-molecule collisional loss measurements suggests that the chemical reaction rates are strongly influenced by the long-range interactions. This observation opens intriguing control possibilities since the long-range interaction can be controlled by selecting quantum states and tuning collision energies via applied electric and magnetic fields.

Reaction rates should be dramatically different if molecules are prepared in a mixture of different hyperfine states as $s$-wave scattering becomes
allowed. We measured the inelastic collision rates for ro-vibronic ground-state molecules that were prepared in a
roughly 50-50 incoherent mixture of the two hyperfine spin states $|-4, 3/2\rangle$ and $|-4, 1/2\rangle$. The time-dependent number
density of trapped molecules was measured for both spin states. We observed the same loss rate for both states, consistent with loss
due to collisions between distinguishable molecules in different spin states. The rate coefficient is determined to be $1.9(4)\cdot
10^{-10}$ cm$^{3}$/s, independent of temperature, as shown in closed squares in Fig.~\ref{mmDecay}(B). In comparison to our measurements
for $p$-wave collisions between spin-polarized molecules, the $s$-wave collision rate between molecules in different hyperfine
states is 10 to 100 times larger for a similar temperature range.

The MQDT model can also be used to estimate collision rates where the dominant collision channel is $s$-wave. Here, the
relevant length scale is determined by the inter-molecular van der Waals potential without any angular momentum barrier~\cite{Kohler2006}.  We assume that when the molecules approach each other within the van der Waals length $\bar{a}$, chemical reactions take place and remove these entrant molecules with a near-unity probability. The universal loss rate coefficient, $\beta = 2(h/\mu)\bar{a}$ \cite{Julienne09a}, predicts a $\beta$ value of $0.8\cdot 10^{-10}$ cm$^{3}$/s, which is a factor of 2 lower than the experimentally observed value.  This difference
suggests that short-range physics may play some role in the loss dynamics. An enhancement in the rate coefficient (up to the energy-dependent unitarity limit, which is $4\cdot 10^{-10}$ cm$^{3}$/s for a gas at 400 nK) is possible if there is a partial reflection of the colliding species
back into the entrance channel \cite{Idziaszek}. The reflected amplitude interferes with the incoming amplitude and can either increase or decrease the rate coefficient from its ``universal" value. Additional theory and experiment are needed to explore this possibility.

In addition to molecule-molecule reactions, the relatively long lifetime of a pure gas of spin-polarized molecules in the optical
trap ($\sim$1 s) affords time to look for chemical reactions between atoms and molecules. The MQDT model described above
can also predict the atom-molecule reaction rates determined by the long-range physics for the universal loss mechanism.
To prepare the atom-molecule mixture, we control the atom density by selectively removing or heating unpaired atoms after
the initial molecule creation \cite{atom}. For these experiments, we typically work with an atom number about 5 to 15
times larger than the molecule number. All atoms and molecules are prepared in their lowest energy states at 545.9 G.
Specifically, K atoms are in their $|F=9/2,\,m_F= -9/2\rangle$ state, Rb in $|F=1,\,m_F=1\rangle$, and KRb in $|-4, 3/2>$.  Here
$F$ is the total atomic spin and $m_F$ is the spin projection. We performed two separate experiments, one with K and
KRb and the second with Rb and KRb, at a  temperature less than 1$\mu$K. In both cases, the background of atoms in the other, undesired, species was less than 1000, corresponding to a density below 5$\times10^9$ cm$^{-3}$. Because both the atoms and the molecules are prepared in
their lowest energy states, trap loss due to inelastic spin-changing collisions is not possible.  However, from Table \ref{potential},
we do expect loss from chemical reactions for the exothermic reaction K + KRb $\rightarrow$ K$_2$ + Rb,
whereas the endothermic Rb + KRb $\rightarrow$ Rb$_2$ + K should be forbidden.

\begin{figure}[!ht]
\begin{center}
\includegraphics[width=\columnwidth]{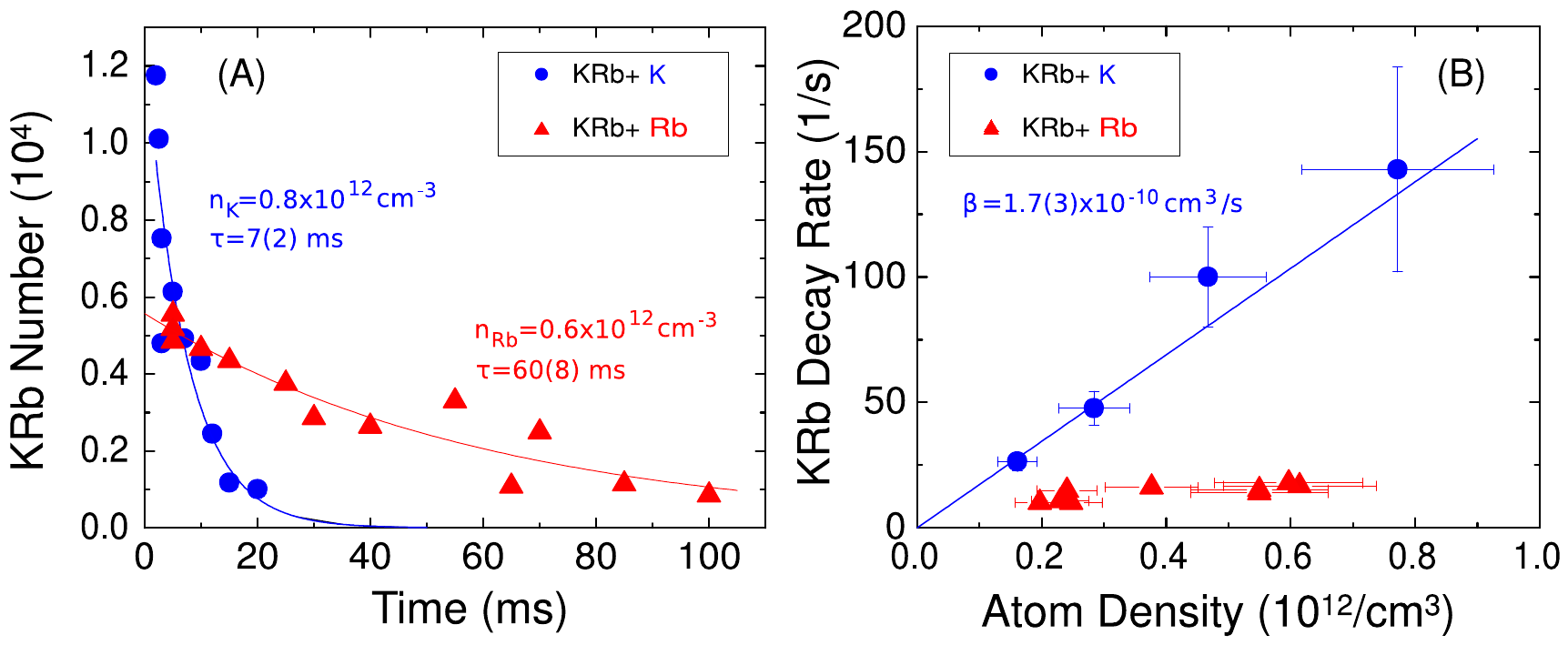}
\caption{\doublespacing Collisions of
atoms and molecules in their lowest energy internal states.
(A) Sample decay curves for the molecular number when subject to collisions with
K atoms (circles) and with Rb atoms (triangles). The atom
numbers are 5 to 15 times higher than the molecule
numbers. The reduced initial molecule number for KRb+Rb
is due to the collisional quenching of the initial weakly bound KRb molecules with Rb~\cite{Zirbel2008a}
before the two-photon Raman process that
transfers the molecules down to the ro-vibronic ground state.
(B) Dependence of the molecule loss rate on the atomic gas density. We observed strong molecule
loss with $\beta=1.7(3)\cdot 10^{-10}$ cm$^{3}/$s for KRb+K collisions and suppressed
loss of KRb for KRb+Rb collisions. The error bars are 1 SD of decay rates and atomic densities.
 \label{am}}
\end{center}
\vspace{-0.05\textwidth}
\end{figure}

For each experiment, we measured the time dependence of the trapped molecule population. Typical molecular loss curves are shown in
Fig.~\ref{am}(A). To extract the inelastic collision rate, we assume that the atom number density is constant. (This is approximately
true because the number of atoms is much larger than the number of molecules.) The trapped molecule number should then decay as
\begin{equation}
\frac{d}{dt} N_{\mathrm{molecule}}=-\beta \cdot N_{\mathrm{molecule}} \cdot
n_{\mathrm{atom}},
\end{equation}
where $N_{\mathrm{molecule}}$ is the number of molecules, $n_{\mathrm{atom}}$ is the atomic density, and $\beta$ is the inelastic rate
coefficient. We can then extract $\beta$ via an exponential fit, $e^{-\beta\cdot n_{\mathrm{atom}}\cdot t}$.  As can be seen in
Fig.~\ref{am}(B), in general we find that the molecules are lost from the trap at a much faster rate when K atoms are present than
when a similar density of Rb is present.

To see if the molecule loss arises from atom-molecule collisions, we measure the dependence of the loss rate on the atom gas density. For
the case of KRb + K, we observe a clear linear dependence on the atomic density and extract an inelastic collision rate coefficient of
$1.7(3)\cdot 10^{-10}$ cm$^{3}/$s. This rate coefficient (due to chemical reactions) can again be predicted from the van der Waals length characterizing the long-range part of the potential between the collision partners. Using similar methods to those used for KRb + KRb, Kotochigova has calculated $C_6$ for KRb + K to be 7020(700) a.~u.~\cite{SvetlanaC6}, which gives a predicted reaction rate of $1.1\cdot 10^{-10}$ cm$^{3}/$s.
Again we find that the MQDT value is close but  somewhat lower than the measured value; this result further solidifies the
importance of long-range quantum scattering in the inelastic process.

In contrast, for KRb + Rb collisions, the density dependence of the loss rate is not obvious (Fig.~\ref{am}(B)). Nevertheless, we again fit the dependence as linear and obtain an upper limit for the rate coefficient of $0.13(4)\cdot 10^{-10}$ cm$^{3}/$s, which
is one order of magnitude smaller than what we measure for KRb + K.

Our measurement is consistent with the fact that for KRb + Rb, there is no two-body chemical reaction pathway.
A possible mechanism for the residual non-zero rate coefficient for KRb + Rb would be collisions of ground-state molecules with undetected molecules in high-lying vibrationally excited states.  These contaminant molecules could be produced by
inelastic collisions of Rb atoms with our weakly bound KRb molecules
\cite{Zirbel2008a} before the two-photon Raman process that produces
ground-state molecules. Another possible loss mechanism is Rb + Rb + KRb three-body inelastic collisions. Further experiments will be needed to check these possibilities.
We note that this suppressed loss rate is observed only when both KRb and
Rb are prepared in their lowest energy internal states at 545.9 G. If either the molecules or the atoms
are in an excited hyperfine state, then the observed inelastic rate coefficient rises again to the order of $10^{-10}$ cm$^{3}/$s. Therefore, for practical reasons,
removing Rb atoms is important for creating a long-lived sample of
ground-state KRb molecules.

Together the studies presented here
show that we have observed barrierless
chemical reactions in the short range, with the rates determined
by long-range scattering dynamics dictated by quantum
statistics, angular momentum barriers, and threshold laws.
We see that a change as seemingly
insignificant as flipping a single nuclear spin
dramatically changes the rate of molecular collisions.
A pure gas of spin-polarized KRb
molecules is long-lived in an optical trap (surviving for a time on
the order of a second) whereas a spin mixed KRb sample decays
10 to 100 times faster.
Our results clearly show that chemical
reactions can proceed with high rates in the ultracold regime.

\bibliography{scibib}

\begin{thebibliography}{10}
\bibitem{removal}
Residual unpaired Rb atoms are removed immediately after creating
weakly bound molecules. We use a sequence that includes adiabatic
rapid passage using frequency-swept microwaves for transitioning Rb atoms
from $|F=1,\,m_F=1\rangle$ to $|2,\, 2\rangle$ and then a pulse of
laser light resonant with the cycling transition from $|2,\,
2\rangle$ to $|F'=3,\,m_F'=3\rangle$ to heat the atoms out of the
trap. The sequence is performed three times within a time span of 1
ms. Unpaired K atoms are removed immediately after the molecules are
transferred into the ro-vibronic ground state. We apply a 0.5 ms
pulse of light on the cycling transition $|F=9/2,\,m_F=-9/2\rangle$
to $|F'=11/2,\,m_F'=-11/2\rangle$. Any trapped atoms left after this
removal are below our detection limit, which is about one thousand
atoms.
\bibitem{KRbC6}
The height of the $p$-wave barrier is determined by the molecule-molecule long-range
potential, namely $\hbar^2 L(L+1)/2\mu R^2-C_6/R^6$, for $L=1$ and
$C_6=16130\,E_h \cdot a_0^6$ (a. u.), where
$E_h=4.36 \cdot 10^{-18}$ J and $a_0=0.53\cdot 10^{-10}$ m.
\bibitem{temp}
We control the molecular gas temperature by varying the initial
temperature of our Rb and K atom gases during forced evaporative cooling.
\bibitem{blackbody}
Loss due to black body radiation driving either vibrational or rotational
excitations is negligible as this is expected to occur on the time scale of 10$^5$ s.
\bibitem{atom}
To remove selective amounts of K,
we partially transfer the population of K atoms from
$|F=9/2,\,m_F=-9/2\rangle$ to $|9/2,\,-5/2\rangle$ using two
successive rf pulses. We then clean out K atoms in the
$|9/2,\,-9/2\rangle$ state using resonant light as in
\cite{removal}. Once these atoms are heated out of the trap, we again use
rf pulses to spin flip all K $|9/2,\,-5/2\rangle$ atoms back down to
$|9/2,\,-9/2\rangle$. To selectively vary the density of Rb, we
perform a complete spin flip from $|1,\,1\rangle$ to $|2,\,2\rangle$
and then turn on resonant light for a duration between 1 $\mu$s and
10 $\mu$s. The resonant light both removes Rb atoms and heats the
remaining gas. After the resonant light pulse, Rb atoms are
transfered back to the lowest hyperfine state $|1,\,1\rangle$. We
then apply an additional resonant light pulse to remove any residual Rb in
$|2,2\rangle$, thus ensuring that all Rb atoms are in the
$|1,1\rangle$ state.
\end{thebibliography}

\bibliographystyle{Science}

\begin{scilastnote}
\item This work is supported by NIST, NSF, and DOE. B. N. acknowledges
support from the NSF and P. S. J. from ONR. We acknowledge S. Kotochigova
for providing her calculations of the van der Waals dispersion coefficients
that are relevant to our experiments.
\end{scilastnote}

\clearpage

\newpage

\clearpage

\newpage

\clearpage

\newpage
\clearpage

\end{document}